\documentclass[preprint]{vgtc}               

\graphicspath{{figures/}{pictures/}{images/}{./}} 

\usepackage{times}                     

\usepackage{tabu}                      
\usepackage{booktabs}                  
\usepackage{lipsum}                    
\usepackage{mwe}                       
\usepackage{amsmath}
\usepackage{amssymb}
\usepackage{amsfonts}

\usepackage[ruled]{algorithm2e}
\SetKwComment{Comment}{/* }{ */}

\usepackage{mathptmx}                  

\onlineid{1211}

\vgtccategory{Research}

\vgtcinsertpkg

\title{Text-based transfer function design for semantic volume rendering}

\author{Sangwon Jeong\thanks{e-mail: sangwon.jeong@vanderbilt.edu}\\ %
        \scriptsize Vanderbilt University %
\and Jixian Li\\ %
     \scriptsize SCI Institute, University of Utah %
\and Christopher Johnson\\ %
     \scriptsize SCI Institute, University of Utah
\and Shusen Liu\\ %
     \scriptsize Lawrence Livermore National Laboratory
\and Matthew Berger\\ %
     {\scriptsize Vanderbilt University}}

\teaser{
  \centering
  \includegraphics[width=\linewidth]{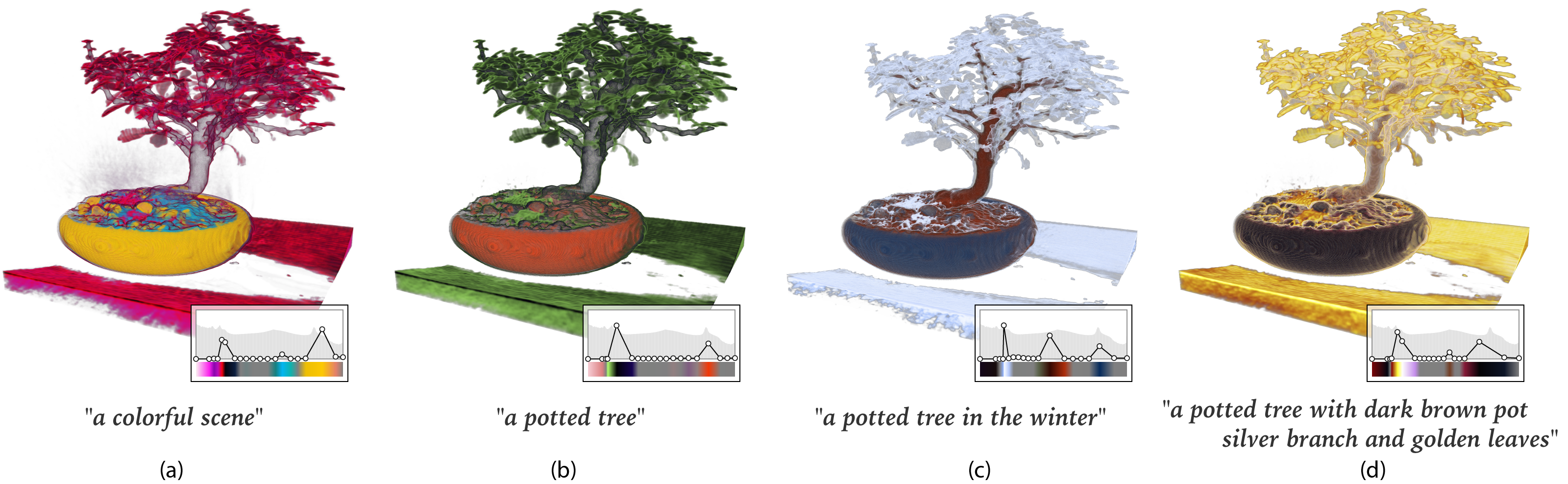}
  \caption{Our method allows users to specify, via natural language, what they wish to see in a volumetric dataset, and in response we optimize for a transfer function that gives volume-rendered images compatible with their description. In this example, we show how (a) a general description permits exploratory analysis of what is contained in a volume, while more precise descriptions lead to TFs that capture the underlying semantics, ranging from (b) natural colors assigned to trees when specifying ``potted tree'', (c) snow in the case of ``winter'', and (d) detailed specification of visual properties being reflected in the volume-rendered image.}
  \label{fig:teaser}
}

\abstract{

Transfer function design is crucial in volume rendering, as it directly influences the visual representation and interpretation of volumetric data. However, creating effective transfer functions that align with users' visual objectives is often challenging due to the complex parameter space and the semantic gap between transfer function values and features of interest within the volume. In this work, we propose a novel approach that leverages recent advancements in language-vision models to bridge this semantic gap. By employing a fully differentiable rendering pipeline and an image-based loss function guided by language descriptions, our method generates transfer functions that yield volume-rendered images closely matching the user's intent. 
We demonstrate the effectiveness of our approach in creating meaningful transfer functions from simple descriptions, empowering users to intuitively express their desired visual outcomes with minimal effort.
This advancement streamlines the transfer function design process and makes volume rendering more accessible to a wider range of users.
} 


\begin{document}

\firstsection{Introduction}
\maketitle

Volume rendering remains foundational to the visual analysis of volumetric scalar fields.
This is in large part due to the interactions offered by volume rendering systems, often taking the form of transfer function (TF) design~\cite{ljung2016state}.
Specifically, users control what they see in the volume by specifying how scalar values map to optical properties, namely opacity and color.
TF design enables a variety of analyses, ranging from open-ended exploration to help understand what exists within a volume, to more detailed analyses that involve exploring specific objects, materials, or features~\cite{laha2015classification}.

However, designing TFs can be a tedious process, particularly for users who have little expertise with volume rendering systems.
In response, numerous methods have been proposed for (semi-) automating the generation of TFs~\cite{maciejewski2009structuring,wang2011efficient,ruiz2011automatic,ma2017volumetric}.
These methods often seek to segment a derived feature space of the volume, for example by using the joint distribution of scalar values and gradient magnitude~\cite{Kindlmann:1999:SAG}, in order to clearly convey well-delineated objects in the rendered image.
Such methods can be effective if the TFs depict structures of interest to the user; otherwise, it becomes necessary for the user to edit the resulting TFs by either refining the opacity or assigning appropriate colors to materials.
Numerous methods have been proposed to make TF design more meaningful~\cite{tzeng2005intelligent,salama2006high,soundararajan2015learning,cheng2018deep}, where users can design TFs with respect to high-level semantics.
Yet these methods often require labeled examples provided by users, e.g. data annotated with pre-defined classes~\cite{salama2006high,cheng2018deep,kim2021image}, or image-based sketching annotations~\cite{tzeng2005intelligent,cheng2018deep}, from which to build statistical models of specific objects.

In this paper, we advocate for more intuitive ways to perform semantic TF design, whereby users can \emph{say} what they want to see in the volume, and a TF that is compatible with their description is produced.
We introduce text-to-transfer function (T2TF), a method for designing TFs via the specification of natural language, namely a text prompt.
T2TF aims to support a variety of analyses common to volume rendering~\cite{laha2015classification}.
For instance, when the user is unclear about what might exist in the volume, they can provide generic descriptions to obtain a general understanding of the data (c.f. Fig.~\ref{fig:teaser}(a)).
From such an overview, a user might wish to be more precise about the objects within a volume, and their visual appearance, and thus they may provide a more detailed description (c.f. Fig.~\ref{fig:teaser}(b)-(d)).
The specification of natural language allows the design of (1) opacity TFs that are intended to represent user-desired structures, and (2) color TFs that represent the visual appearance of such structures.
For instance, when specifying the season of winter for depicting a tree in a given volume, we obtain faded white/brown colors of a tree, reflecting the characteristics of winter (c.f. Fig.~\ref{fig:teaser}(c)).

In order to find a TF that satisfies a user's description, we combine vision-language models, namely CLIP~\cite{radford2021learning}, with differentiable volume rendering~\cite{weiss2021differentiable} to optimize for TFs.
The most straightforward approach to the problem is to find a TF such that the CLIP-based similarity between the volume-rendered image, and the user's description, is maximized.
However, this is a highly under-constrained optimization problem, as \emph{many} possible TFs, and volume-rendered images, can give high alignment with the description, as measured by CLIP.
T2TF addresses these challenges by (1) better constraining optimization, (2) permitting control on the density of rendered materials, and (3) ensuring that the found TFs are interpretable.
In detail, inspired by contrastive representation learning~\cite{chen2020simple,radford2021learning}, our optimization objective ensures that volume-rendered images are well-aligned with the user's description, while remaining poorly aligned with random descriptions.
Secondly, in order to control for the density of materials depicted in volume-rendered images, we incorporate a ray-based prior on transmittance~\cite{Lombardi:2019}.
Last, we introduce a regularization on TFs to minimize their complexity, namely that TFs should only be as complex as necessary.
This regularization is motivated by the intended use case of our method, where the output TF should be editable by the user, and thus must be easy to understand.

We demonstrate the benefits of T2TF on a variety of volumes, spanning a variety of text descriptions.
We show how descriptions that reference different objects in a volume can lead to TFs that faithfully identify the relevant objects, as largely indicated by opacity (c.f. Fig.~\ref{fig:visiblemale}).
Moreover, we show how color TFs tend to reflect the semantics of user descriptions, all achieved without having to explicitly mention colors (c.f. Fig.~\ref{fig:teaser}(b)).
We further show how our method can support analysis at varying levels of detail, ranging from the specification of general prompts that are intended to give an overview, to more precise prompts intended to depict detailed information within volumes.

In summary, our contributions consist of: (1) an approach to semantics-based TF design based on the specification of natural language, (2) a means of controlling material density in optimized TFs, (3) ensuring TFs are of minimum complexity, thus facilitating post-optimization editing of TFs, and (4) the generality of our method across a large number of volumes and text descriptions.

\section{Background}

The problem we address in this paper can be summarized as follows: given a volumetric scalar field, and a text-based description of the volume that is of interest to the user, the objective is to find a TF whose corresponding volume-rendered images can be accurately described by the user’s text.
To solve this problem, T2TF draws from two main research areas: volume rendering, and vision-language models (VLM).
To help understand the details of our approach, we first provide the necessary background on both areas.

\paragraph{Volume rendering and TF optimization} Our method relies on the absorption-plus-emission model of volume rendering~\cite{Max:1995:OMF}.
Specifically, given a scalar field $f : \mathbb{R}^3 \rightarrow \mathbb{R}$ and a viewing ray $\mathbf{v}(t)$ with length parameter $t$, for which the entry point of the volume is $t = 0$ and exit point $t = T$, the volume rendering equation amounts to:
\begin{equation}
    \mathbf{C}(\mathbf{v} ; \phi) = \int_{0}^T T\left( \mathbf{v},t ; \phi \right) \sigma_{\phi}(f_{\mathbf{v}}(t)) \mathbf{c}_{\phi}(f_{\mathbf{v}}(t)) d t,
    \label{eq:vrint}
\end{equation}
where $f_{\mathbf{v}}(t) = f(\mathbf{v}(t))$ is the value of the scalar field evaluated at the location $\mathbf{v}(t)$ along the ray, and the functions $\sigma_{\phi}$ and $\mathbf{c}_{\phi}$ are respectively the density and color TFs, namely 1D functions that depend on parameters $\phi$.
The function $T$ represents transmittance, or the probability that light along a ray traversing the volume from entry (0) to some current distance ($t$) will not be absorbed by light:
\begin{equation}
T\left( \mathbf{v}, t ; \phi \right) = \exp \left( -\int_0^t \sigma_{\phi}(f_{\mathbf{v}}(u)) du \right)
    \label{eq:Tint}
\end{equation}
Standard numerical integration to approximate the integral~\cite{Max:1995:OMF} leads to the following iterative expressions for computing color and transmittance:
\begin{align}
& \mathbf{C}(\mathbf{v} ; \phi) \approx \sum_{n=1}^N T_n(\mathbf{v} ; \phi) \left(1 - \exp(-\sigma_{\phi}(f_{\mathbf{v}}(t_n)) \delta)\right) \mathbf{c}_{\phi}(f_{\mathbf{v}}(t_n)), \label{eq:vr} \\
& T_n(\mathbf{v} ; \phi) = \prod_{i=1}^n \exp\left(-\sigma_{\phi}(f_{\mathbf{v}}(t_i)) \delta \right), \label{eq:T}
\end{align}
where $N$ represents the number of steps taken along the ray, $t_n$ the ray distance for step $n$, and $\delta$ the step size

Notably, if density and color TF access are both differentiable functions, then the color in Eq.~\ref{eq:vr} can be viewed as a differentiable function in the TF (namely, its parameters $\phi$), and camera pose.
This has led to numerous methods for finding rendering parameters via gradient-based optimization of volume rendering.
For instance, Weiss et al.~\cite{weiss2021differentiable} optimize TFs to match an image reference, further extended towards the comparison of fields~\cite{neuhauser2023transfer}, as well as the exploration of a collection of TFs~\cite{pan2023differentiable}.
For T2TF, however, we do not have a reference image on which to compute an image-based loss.
Rather, given a text description, we require a means of scoring how well a volume-rendered image matches the user's text.



\paragraph{Vision-language models} Recently, a number of foundation models have been developed to align images with text~\cite{radford2021learning, bao2022vlmo, wang2024visionllm}.
Specifically, the model of CLIP~\cite{radford2021learning} has shown that when trained on a massive amount of data, namely hundreds of millions to billions~\cite{schuhmann2022laion} of captioned images, high-capacity neural networks are capable of accurately scoring whether a given image is well-described by a text prompt.
This is beneficial for T2TF in the following way: given a volume-rendered image denoted $\mathbf{I}$, and a user-provided text description $y$, we can compute a vector-based encoding of both the image $\mathbf{e}_I(\mathbf{I})$ and text $\mathbf{e}_y(y)$, and their dot product $\mathbf{e}_I(\mathbf{I})^T \mathbf{e}_y(y)$ gives us a score of image-text alignment.
At a basic level, T2TF seeks a TF such that its corresponding volume-rendered image yields a high CLIP score with the user's text.

A major motivation for us in using CLIP is its strong ability to perform zero-shot categorization of images~\cite{novack2023chils}, demonstrating a deep understanding of both visual features and natural language~\cite{dai2022enabling}.
However, despite its robustness, CLIP nevertheless gives reduced performance under distribution shift~\cite{tu2024closer}.
Specific to our problem setting, we anticipate covariate shift due to the simplified illumination model (c.f. Eq.~\ref{eq:vr}), for instance, the rendered tree in Fig.~\ref{fig:teaser} appears different from an actual tree.
Handling such shortcomings has been addressed in the computer vision community, namely for the purposes of generating 3D radiance fields~\cite{jain2022zero,poole2022dreamfusion,lee2022understanding,wang2023score} solely from a given text description.
In particular, our work is motivated by prior work~\cite{jain2022zero,poole2022dreamfusion,wang2023score}, in which a radiance field, e.g. a neural network that reports density and color at each location, is optimized such that its volume-rendered images either give high CLIP scores~\cite{jain2022zero}, or high likelihood with respect to a diffusion model~\cite{poole2022dreamfusion,wang2023score}.
Nevertheless, a distinction in our work is that we do not have the same degree of flexibility in what is being optimized.
Specifically, the 3D field is fixed, rather than learned, as we instead aim to find the parameters of a TF for which to produce density and color.
Thus, unique to our setting, there are several considerations that inform our method.
First, we require a CLIP-based objective that can mitigate shortcomings in distribution shift. 
Secondly, a user's description might be imprecise, and thus distribution shift notwithstanding, a variety of TFs can produce volume-rendered images that are all in agreement with the description.
Rather than leave the generated TF to chance, we wish to introduce control over the rendering process where feasible, namely the level of density in rendered materials, while also ensuring that TFs are only as complex as they need to be, and thus easily editable.

\section{Text-based transfer function design}

\begin{algorithm}[!t]
\caption{Text-to-transfer function} \label{algorithm}

\SetKwInOut{KwIn}{Input}
\SetKwInOut{KwOut}{Output}

\KwIn{Scalar field $f$, text description $y$}
\KwOut{Parameters $\phi$ for density TF $\sigma_{\phi}$ and color TF $\mathbf{c}_{\phi}$}

$\mathbf{c}_{\phi}$ : random initialization, $\sigma_{\phi} : ave(T_N(\mathbf{V}_0 ; \phi)) \approx \rho$\;

\For{$i \gets 1$ \KwTo $N$}
{

$\mathbf{V} \sim \mathcal{V}$ (sample camera pose)\;
$\mathbf{B} \sim \mathcal{B}$ (sample background image)\;
$\mathbf{\check{y}} \sim \mathcal{P}$ (sample negative prompts)\;

$\mathbf{I} \gets \mathbf{C}(\mathbf{V} ; \phi) + T_N(\mathbf{V} ; \phi) \mathbf{B}$\;
$\mathcal{L}_{CLIP}(\phi) \gets CE\left( \mathbf{e}_I(\mathbf{I})^T \mathbf{e}_y(y) ; \mathbf{e}_I(\mathbf{I})^T \mathbf{e}_y(\mathbf{\check{y}}) \right)$\;
$\mathcal{L}_{density}(\phi) \gets - \log P_{beta}(T_N(\mathbf{V} ; \phi) ; a, b)$\;
$\mathcal{L}_{reg}(\phi) \gets \lambda_1 \lVert \sigma_{\phi} \rVert_1 + \lambda_2 \lVert \mathbf{c}_{\phi} - 0.5 \rVert^2_2$\;
$\phi \gets \phi - \eta \nabla_{\phi}(\mathcal{L}_{CLIP}(\phi) + \mathcal{L}_{density}(\phi) + \mathcal{L}_{reg}(\phi))$\;

}

\end{algorithm}

\begin{figure}[!t]
    \centering
    \vspace{-2mm}
    \includegraphics[width=0.98\linewidth]{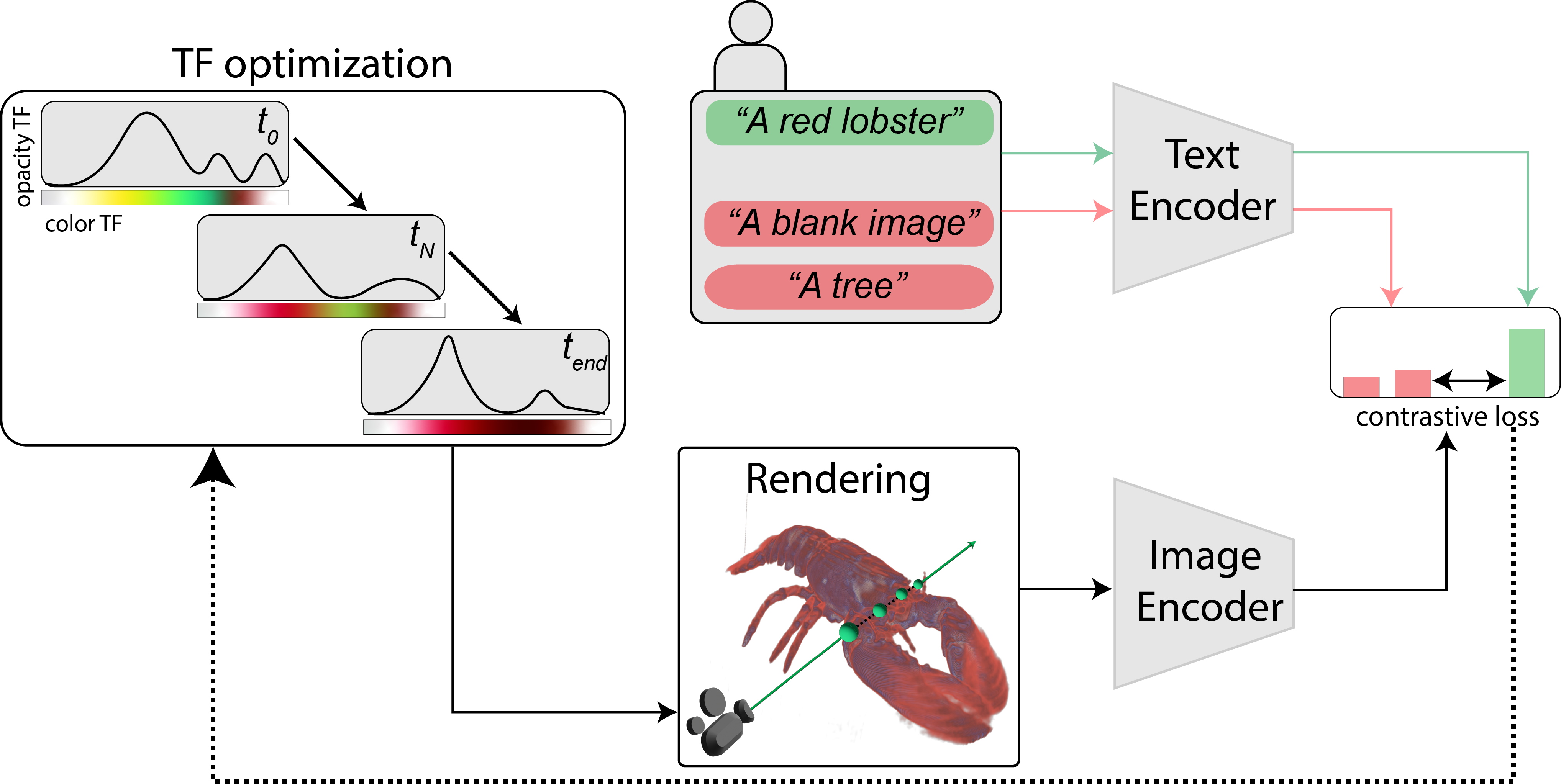}
    \caption{We show an overview of T2TF. Given a TF, we volume render from a camera view to produce an image and subsequently obtain a CLIP-based encoding of the image. We then compare the image with a CLIP-based encoding of the user's prompt, expressive of their visualization objective, with negative sampled prompts. We form a contrastive loss between the user's prompt and negative prompts, giving us a way to update the TF via gradient-based optimization.
    }
    \vspace{-2mm}
    \label{fig:overview}
\end{figure}

In this section we describe our method for mapping text descriptions to TFs, please see Fig.~\ref{fig:overview} for a graphical overview, and Algorithm~\ref{algorithm} for a description of our optimization procedure.
We defer additional details on optimization to supplementary material.

\paragraph{TF representation}

We choose to represent both density ($\sigma_{\phi}$) and color ($\mathbf{c}_{\phi}$) TFs as piecewise linear functions.
Thus, TF parameters $\phi$ are comprised of (1) control points, restricted to the range of $f$, and for each control point (2) a nonnegative value representing density, and (3) a 3-vector representing the color in RGB space, each entry constrained to $[0, 1]$.
We chose this representation due to its predominance in scientific visualization software~\cite{ayachit2015paraview,childs2012visit}, and its ease of editing TFs via dragging control points, post-optimization.

\paragraph{Objective} 

Our goal is to find a TF that satisfies the user's description in the volume-rendered output, while also allowing for control on the density of rendered materials, and last, ensuring the found TF is of low complexity.
Our objective thus follows as a combination of three loss terms addressing each of these considerations, wherein we aim to minimize:
\begin{equation}
\mathcal{L}(\phi) := \mathcal{L}_{CLIP}(\mathbf{I}(\mathbf{V}, \mathbf{B}; \phi)) + \mathcal{L}_{density}(T_N(\mathbf{V} ; \phi)) + \mathcal{L}_{reg}(\phi) \label{eq:full_loss}.
\end{equation}
Specifically, for a given camera pose $\mathbf{V}$ and background image $\mathbf{B}$, we denote $\mathbf{I}(\mathbf{V}, \mathbf{B}; \phi)$ as the volume-rendered image.
The term $\mathcal{L}_{CLIP}$ represents a CLIP-based loss, intended to align the image with the user's description. 
Denoting $T_N(\mathbf{V}; \phi)$ as the transmittance for each viewing ray associated with pose $\mathbf{V}$, the term $\mathcal{L}_{density}$ controls for material density via placing a prior on transmittance.
Last, the term $\mathcal{L}_{reg}(\phi)$ is a complexity penalty on the TF.
We optimize over a distribution of camera poses $\mathbf{V} \sim \mathcal{V}$, and background images $\mathbf{B} \sim \mathcal{B}$, in order to ensure that the user's description is satisfied under multiple views, while also ensuring that CLIP does not overfit to the background, following prior work~\cite{jain2022zero,poole2022dreamfusion}.
Prior to optimization, the color TF is randomly initialized, while the density TF is found such that average transmittance, at a fixed view $\mathbf{V}_0$, approximately satisfies a specified value $\rho$.
In practice $\rho = .05$, to ensure the field's full range of values contributes to volume rendering.

\paragraph{CLIP-based learning of TFs}

A straightforward way to find TFs is to simply maximize the CLIP score of a volume-rendered image with the user's description.
However, we found this to be often unreliable, emphasizing transparent materials and volume background; please see supplemental for comparisons.
As it is possible for the given image to satisfy \emph{many} different descriptions, we aim to find a TF whose volume-rendered images are \emph{unique} to the user's description.
To this end, we propose a contrastive loss~\cite{chen2020simple,radford2021learning} for learning TFs.
Specifically, at each step of optimization, we sample, uniformly at random, a collection of \emph{negative} prompts from a predefined set of image-captions $\mathcal{P}$ -- in practice, we use LAION~\cite{schuhmann2022laion}.
We then compute the CLIP score between the volume-rendered image, and (1) the user's prompt $y$, as well as (2) the collection of sampled negative prompts denoted $\mathbf{\check{y}} \sim \mathcal{P}$.
We treat these scores as logits, subsequently passed to a softmax to form a discrete distribution, and then compute a cross-entropy loss to discriminate a user's description from negative prompts.

\begin{figure*}[!hbt]
    \centering
    \includegraphics[width=\textwidth]{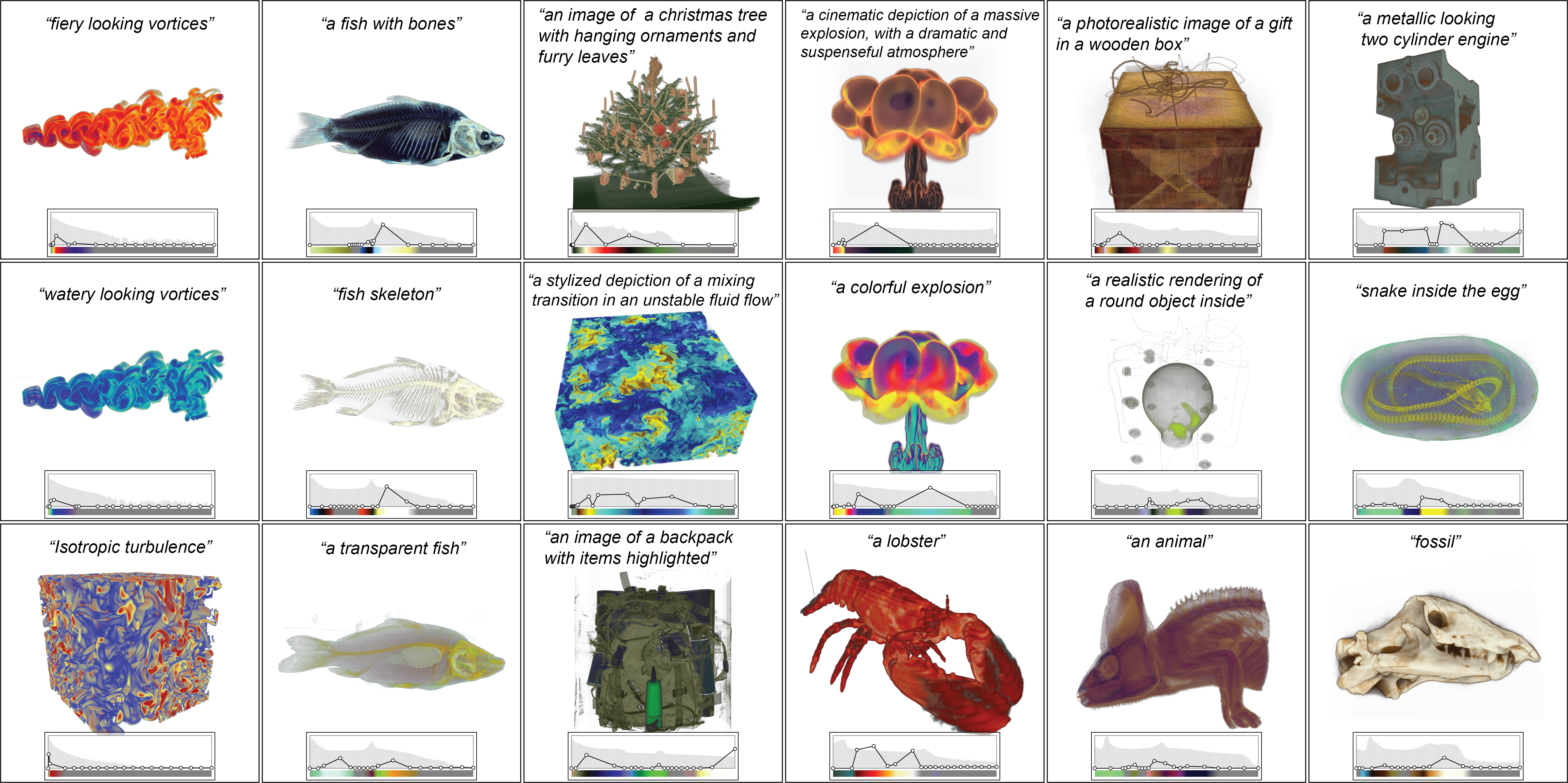}
    \caption{T2TF results for a variety of volumes (scanned data, numerical simulations) and prompts (general vs. detailed descriptions).}
    \label{fig:gallery}
\end{figure*}

We can further extend our formulation to help the user \emph{suppress} certain structures in the output; this is most useful when it is challenging to include these omissions in the given description.
Specifically, when we sample negative prompts $\mathbf{\check{y}}$, we append the user's supplied set of negative prompts.
As shown in Fig.~\ref{fig:visiblemale}(d), user-supplied negative prompting can help omit undesirable materials, for instance skin, skull, to capture just one material, namely teeth.

\paragraph{Density prior}

In controlling for material properties, not everything can be so easily spoken; this is especially the case for material density, which is continuous, and thus difficult to express in language.
To this end, following Lombardi et al.~\cite{Lombardi:2019}, we place a beta prior on transmittance $T_N(\mathbf{V}; \phi)$, with prior parameters $a = b < 1$ in order to give a bias towards the transmittance being either 0 or 1.
This is helpful in biasing discovered materials of the TF to have higher density, while also discouraging semi-transparency, as viewing rays whose transmittance is small are biased towards zero transmittance.
We found this prior to give consistent results, in comparison to other methods considered for neural radiance fields~\cite{jain2022zero,kim2022infonerf}; please see supplemental for comparisons.

\paragraph{TF regularization}

Although the TF that we find will have high density placed in relevant portions of the volume's range space, unless appropriately regularized, it might also place high density in portions that have little impact on the volume-rendered results, or occluded by the relevant material.
If one wants to edit the resulting TF, this can lead to some confusion about what is relevant to edit.
To this end, we aim to minimize the complexity of the TF, wherein we apply a sparsity-promoting $l_1$ regularization to the density evaluated at the control points.
This will encourage low-density assignment to function values not relevant to the materials displayed.
Moreover, for the color TF, we additionally apply an $l_2$ regularization towards gray colors; this is done primarily for optimization stability purposes, so that colors do not saturate at the extremes (e.g. 0 or 1) over the course of optimization.

\section{Results}
\begin{figure}[htb]
    \centering
    \includegraphics[width=\columnwidth]{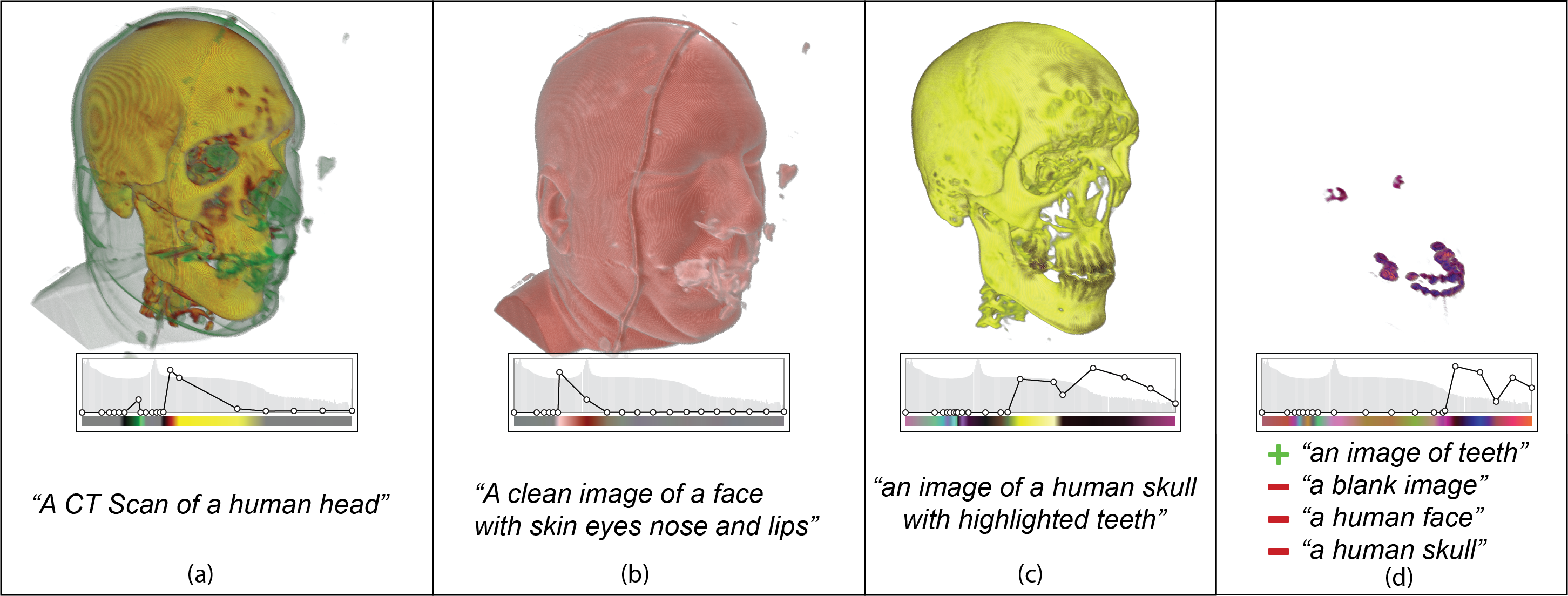}
    \caption{Volume-rendered images of the visible male dataset. In (a), the generic prompt results in the transfer functions that show an overview of features in the volume. Then, in (b)-(d), different opacity transfer functions highlight the skin and skull. This example illustrates the capability of T2TF to adjust opacity transfer functions}
    \label{fig:visiblemale}
    \vspace{-2mm}
\end{figure}
This section demonstrates the images produced from T2TF transfer functions on multiple datasets. The datasets are from Open Scientific Visualization Datasets \footnote{https://klacansky.com/open-scivis-datasets/} through OpenVisus \footnote{https://visus.org/}. Details of each dataset are provided in the supplementary material. 


We have demonstrated in Fig.~\ref{fig:teaser} that T2TF can produce different color TFs that match the theme of the prompt.
In Fig.~\ref{fig:visiblemale}, T2TF is asked to create images highlighting various features of the human head from the visible male dataset, a CT scan of a human head.
Fig.~\ref{fig:visiblemale}(a) results from a generic prompt asking for a CT scan of a human head.
The found TF gives an overview of the scene, depicting two large features: a transparent skin and a solid skull, shown as two peaks in the opacity TF.
For a more precise description, in Fig.~\ref{fig:visiblemale}(b) we request the skin of the scan.
We use the word ``clean'' to reduce ambient noise, which has similar scalar values to the skin, and we find the single peak opacity TF effectively renders the skin.
In Fig.~\ref{fig:visiblemale}(c) request an image of both the skull and the teeth, where we find the resulting TF contains two peaks in the opacity TF.
These peaks correspond to the scalar values associated with the skull and teeth, while other regions are mostly flat.
The color TF uses a contrasting color to the skull, in order to highlight the teeth.
Fig~\ref{fig:visiblemale}(d) shows our setup to highlight one small feature, the teeth, of the CT scan.
We found it is most effective to add negative prompts to remove the large feature when we want to isolate smaller features in the scene.
To this end, we provide three additional negative prompts: ``a human skull'' and ``a human face'', in order to suppress these features, and ``a blank image'', to ensure the TF does not degenerate to a trivial solution.
As shown, the opacity TF only peaks at the scalar value representing the teeth in the dataset.

In Figure \ref{fig:gallery}, we showcase our approach by providing an assortment of T2TF-generated transfer functions and their volume-rendered images. The results suggest that T2TF can produce meaningful TFs for a diverse set of volumetric data, ranging from scanned data to numerical simulations, and different types of prompts, from general to more detailed descriptions of scenes.

\section{Discussion}

In this work, we have introduced T2TF, a method for designing TFs via the specification of natural language.
Although our results show promise for this new way of semantically specifying TFs, we acknowledge several limitations of our approach.
First, the computational cost required to optimize for a TF places our method at far from interactive.
The complexity is largely dominated by the cost to backpropagate through the volume renderer, and this further depends on the size of the volume.
Specifically, for the visible male volume, it takes approximately $0.5$s for a single step of gradient descent for a single camera pose, while for the Rayleigh-Taylor instability volume, this takes around $2$s per step.
For our results, we conservatively set the number of optimization steps in order to ensure convergence, namely 300 steps, but in practice, we find that the number of steps can be reduced while still obtaining similar results; future work will investigate trade-offs between speed and quality.
Secondly, we assume that what the user specifies in their description is visible at arbitrary camera poses.
However, for features that are more localized, and thus view-specific, this assumption may no longer hold.
A method for viewpoint optimization, similar to Weiss et al.~\cite{weiss2021differentiable}, is therefore required to address this issue.



Our work opens doors to many future directions.
In many ways, our work is complementary to the extensive body of research in volume rendering, and thus we intend to explore how multidimensional TFs~\cite{kniss2001interactive,KNISS2005189}, illumination models~\cite{hernell2009local}, and image-based TF editing~\cite{guo2011wysiwyg} can enhance T2TF.
Moreover, we need not limit ourselves to CLIP, and can instead use proper likelihood-based models, namely text-conditional diffusion models~\cite{poole2022dreamfusion,wang2023score}.
More broadly, we will investigate multi-modal large language models~\cite{liu2024visual} for T2TF, as they demonstrate stronger language understanding and better vision-language alignment.
In addition, we plan to improve the performance of our method to minimize the latency between users providing prompts, and users obtaining their desired TFs.

\acknowledgments{
This work was performed under the auspices of the U.S. Department of Energy by Lawrence Livermore National Laboratory under Contract DE-AC52-07NA27344. The project is partly supported by LLNL LDRD (23-ERD-029).
This work is reviewed and released under LLNL-CONF-863784. This work was partially supported by the Intel OneAPI CoE, the Intel Graphics and Visualization Institutes of XeLLENCE, and the DOE Ab-initio Visualization for Innovative Science (AIVIS) grant 2428225.
}

\bibliographystyle{abbrv-doi}

\bibliography{template}
\clearpage

\newpage
\newpage
\appendix

\setcounter{page}{1}

\section{Optimization details}

In this section, we provide additional details on our optimization procedure.

\subsection{TF parameters}

Recall that the TF parameters $\phi$ consist of control point locations, and at each location a value of density and color.
We represent the control points in a relative manner, via the spacing between adjacent control points.
To ensure strictly nonnegative values we first transform these unconstrained parameters via the softplus function.
We then compute their cumulative sum to arrive at absolute positions, and subsequently normalize the locations such that the first control point is assigned the minimum value of the scalar field, and the last control point is assigned the maximum value of the scalar field.
Both density and color parameters take on unconstrained values, where the density is subsequently mapped to the range $[0, 255]$ via a hyperbolic tangent nonlinearity, and likewise, each RGB color channel is mapped to $[0, 1]$ via hyperbolic tangent.
We found constraining density to a particular range easier to optimize, in contrast with allowing density to take on arbitrarily-large nonnegative values.

We randomly initialize the values of the color TF to be in the range $[0.3, 0.7]$; this is done to ensure that the color is not arbitrarily close to 0/1, thus avoiding saturated gradients during optimization.
The density parameters are initialized such that for a fixed viewpoint $\mathbf{V}_0$, the transmittance $T_N$, averaged over all viewing rays, is approximately a specified value $\rho = .05$.
Density parameters are found via gradient descent, where we initialize the density such that it is inversely proportional to the estimated density of the scalar field, approximated through the volume histogram.

\subsection{Optimization parameters}

In optimizing for TF parameters, we use stochastic gradient descent (SGD) with momentum, with the learning rate set to $10$, and momentum set to $0.75$.
We find that more sophisticated optimization schemes, for instance, Adam, can lead to saturated gradients for the values at individual control points, be it color or density.
We run SGD for a total of 300 steps, and employ a linear learning rate annealing.
Taking a single gradient step involves sampling a fixed number of random camera poses, performing volume rendering for each view, evaluating the loss function (c.f. Eq.~\ref{eq:full_loss}), and averaging the resulting per-view gradients.
In practice we find that taking 3 camera poses at each step gives improved robustness compared to just a single view, but at the expense of additional computation.
Camera poses are sampled such that we allow for arbitrary camera yaw angles to orbit the volume, restrict pitch angle to be in the range $[-\frac{\pi}{14}, \frac{\pi}{14}]$, and sample a distance to the volume such that the camera is at least twice the distance from the center of the volume, and at most four times the distance from the center.

We find that a careful choice of background image on which to composite the volume-rendered image can help mitigate limitations in the CLIP score, and prevent CLIP from overfitting to just a single, constant, background.
Thus, we leverage the background augmentation strategy in Jain et al.~\cite{jain2022zero}, whereby a background of either a checkerboard pattern, a noise pattern, or a random Fourier feature is chosen uniformly at random.
In practice, however, we find that such background augmentation can lead to instability in optimization in the first few steps of gradient descent.
Thus, for the first 25 steps of optimization we choose backgrounds of constant color, namely a random shade of gray.
Afterwards, we switch to background augmentation for the rest of optimization.
Moreover, we only the density prior beginning at step 100 of optimization; this is done to ensure that the general scene is well-represented, and thus the prior serves to refine the scene towards a particular density bias.

The weight for the density regularization $\lambda_1$ is set to $2 \cdot 10^{-5}$, while the weight for the color regularization is set to $8 \cdot 10^{-4}$.
For our contrastive loss, at each optimization step we sample $128$ prompts, uniformly at random, from $12$M prompts subsampled from the LAION dataset~\cite{schuhmann2022laion}.

\section{Datasets}
The majority of datasets used to demonstrate T2TF are from Open Scientific Visualization Datasets at \url{https://klacansky.com/open-scivis-datasets/} through OpenVisus (\url{https://visus.org/}). We listed the datasets we used in \autoref{tab:datasets}. Additionally, we used the ``Half Cylinder Ensemble'' flow simulation by Rojo and G{\"u}nther \cite{BaezaRojo19SciVisa}.

\begin{table*}[!t]
    \centering
    \begin{tabular}{|c|c|c|}
    \hline
         Dataset Name& Resolution & Credit to   \\\hline\hline
         Bonsai& $256\times 256\times256$& volvis.org and S. Roettger, VIS, University of Stuttgart \\\hline
         Head (Visible Male) & $128\times 256 \times 256$ & National Library of Medicine, National Institutes of Health, USA  \\\hline
         Isotropic Turbulence & $ 256 \times 256\times 256$&  \begin{tabular}{@{}c@{}}Dataset provided Gregory D. Abram and Gregory P. Johnson,\\ Texas Advanced Computing Center, The University of Texas at Austin. \\ Simulation by Diego A. Donzis, Texas A\&M University,\\ P.K. Yeung, Georgia Tech.\end{tabular}   \\\hline
         Carp & $256\times256\times512$  & Michael Scheuring, Computer Graphics Group, University of Erlangen, Germany  \\\hline
         Christmas Tree & $512\times 499\times 512$ & Armin Kanitsar et al. 2002 \cite{christmas_tree} \\\hline
         Rayleigh-Taylor Instability & \begin{tabular}{@{}c@{}}$1024 \times 1024 \times 1024 $ \\ cropped to \\ $1024\times 1024\times 512$\end{tabular} & Andrew W. Cook, William Cabot, and Paul L. Miller \cite{miranda}\\\hline
         Backpack & $512\times 512\times 373$ &volvis.org and Kevin Kreeger, Viatronix Inc., USA \\\hline
         CSAFE Heptane Gas & $302\times 302\times 302$ &The University of Utah Center for the Simulation of Accidental Fires and Explosions. \\\hline
         Lobster & $301 \times 324\times 56$ &volvis.org and VolVis distribution of SUNY Stony Brook, NY, USA\\\hline
         Christmas Present & $492\times 492\times 442$ & Christoph Heinzl, 2006\\\hline
         Chameleon & 
         \begin{tabular}{@{}c@{}}$1024 \times 1024 \times 1080 $ \\ downsampled to \\ $512\times 512\times 540$\end{tabular}
         & Chamaeleo calyptratus. Digital Morphology, 2003.\\\hline
         Engine & $256\times 256\times 128$ & volvis.org and General Electric \\\hline
         Kingsnake & \begin{tabular}{@{}c@{}}$1024 \times 1024 \times 795 $ \\ downsampled to \\ $512\times 512\times 398$\end{tabular} & \begin{tabular}{@{}c@{}}DigiMorph.org, The University of Texas High-Resolution X-ray CT Facility (UTCT),\\ and NSF grant IIS-9874781\end{tabular}\\\hline
         Spathorhynchus Fossorium&
         \begin{tabular}{@{}c@{}}$1024 \times 1024 \times 750 $ \\ downsampled to \\ $512\times 512\times 375$\end{tabular} &Matthew Colbert, 17 February 2005 \\\hline
         Frog & $256\times 256\times 44$ & Lawrence Berkeley Laboratory, USA \\\hline
         
    \end{tabular}
    \caption{Datasets from Open Scientific Visualization Datasets}
    \label{tab:datasets}
\end{table*}

\subsection{CLIP loss}
When training the T2TF model, maximizing the CLIP score alone sometimes leads to undesired outcomes. We utilize the contrastive loss in T2TF to optimize for user-provided prompts while discouraging the results that lead to negative prompts. This is done to ensure the resulting TF can generate images that better match desired prompts.

\begin{figure}[t]
    \centering
    \includegraphics[width=\columnwidth]{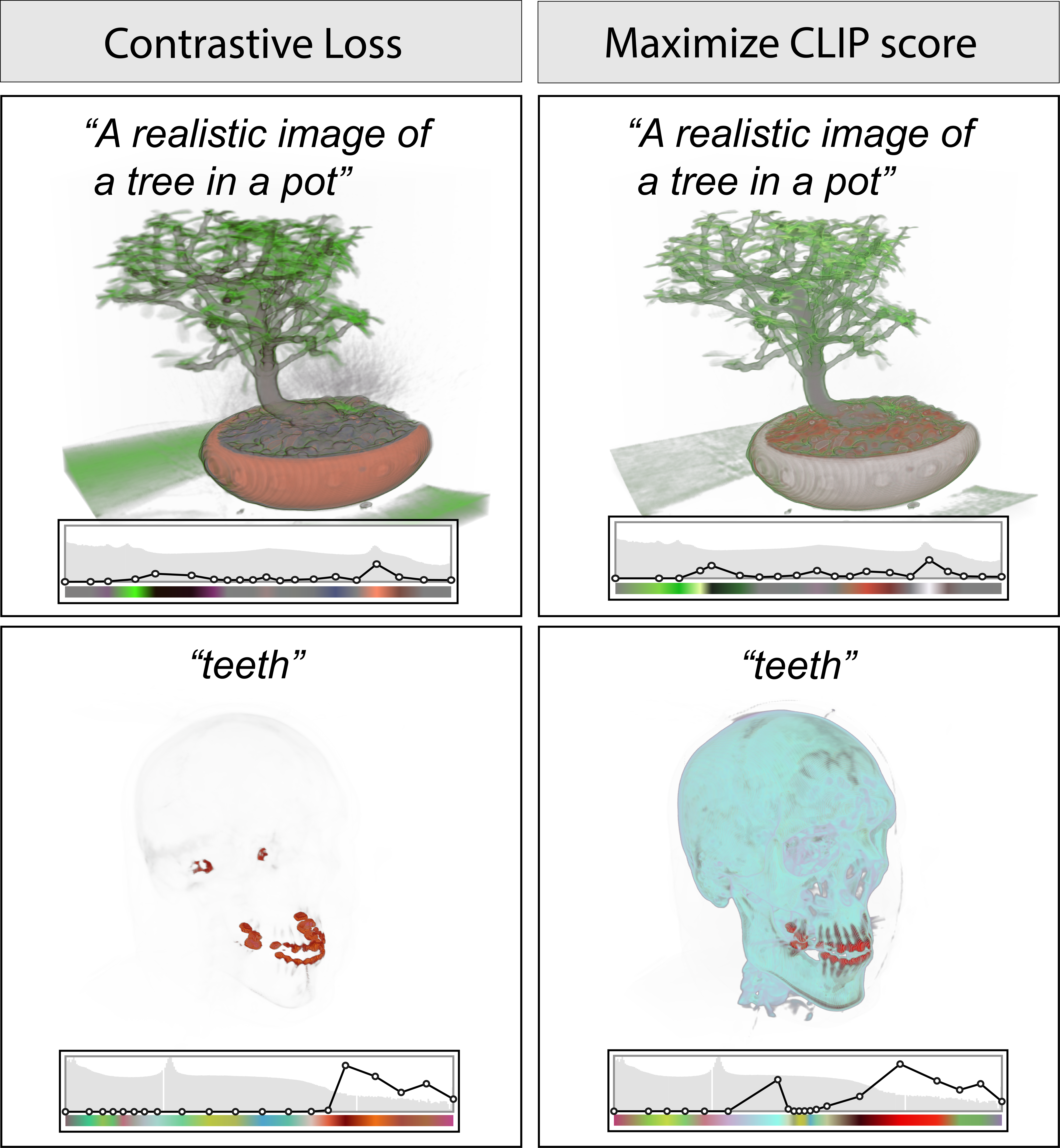}
    \caption{Transfer functions and images comparing the T2TF with contrastive loss and maximizing CLIP score. With contrastive loss, the resulting transfer function can produce images closer to desired prompts.}
    \label{fig:cliploss-comparison}
\end{figure}

Fig.~\ref{fig:cliploss-comparison} shows two pairs of example usages of the contrastive loss compared to maximizing CLIP score. The first row of Fig.~\ref{fig:cliploss-comparison} shows that using contrastive loss results in a TF that gives a more dense, realistic rendering as opposed maximizing only the CLIP score. Here, we use randomzied negative prompts from LAION\cite{schuhmann2022laion}. We intend to train against random prompts so the resulting image can better match the given user prompt. The resulting image using contrastive loss shows the tree is closer to the natural color and thickness of bonsai trees whereas using only the CLIP score leads to a more transparent, unnatural look. In most of our experiments, we use ``a blank image'' and similar phrases as the negative prompts to push for some results. 

Utilizing negative prompts is also effective in isolating features from medical images. In the second row of Fig.~\ref{fig:cliploss-comparison} we show that with or without contrastive loss, T2TF can find transfer functions that highlight the teeth. However, contrary to randomzied negative prompts we use a curated set of negative prompt in this exmaple. On the left, we specified ``a human face'' and ``a human skull'' as the negative prompts. As a result, the optimized opacity transfer function now only peaks at the value of teeth.



\subsection{Comparison of Density Prior}
\begin{figure}[!t]
    \centering
    \includegraphics[width=\columnwidth]{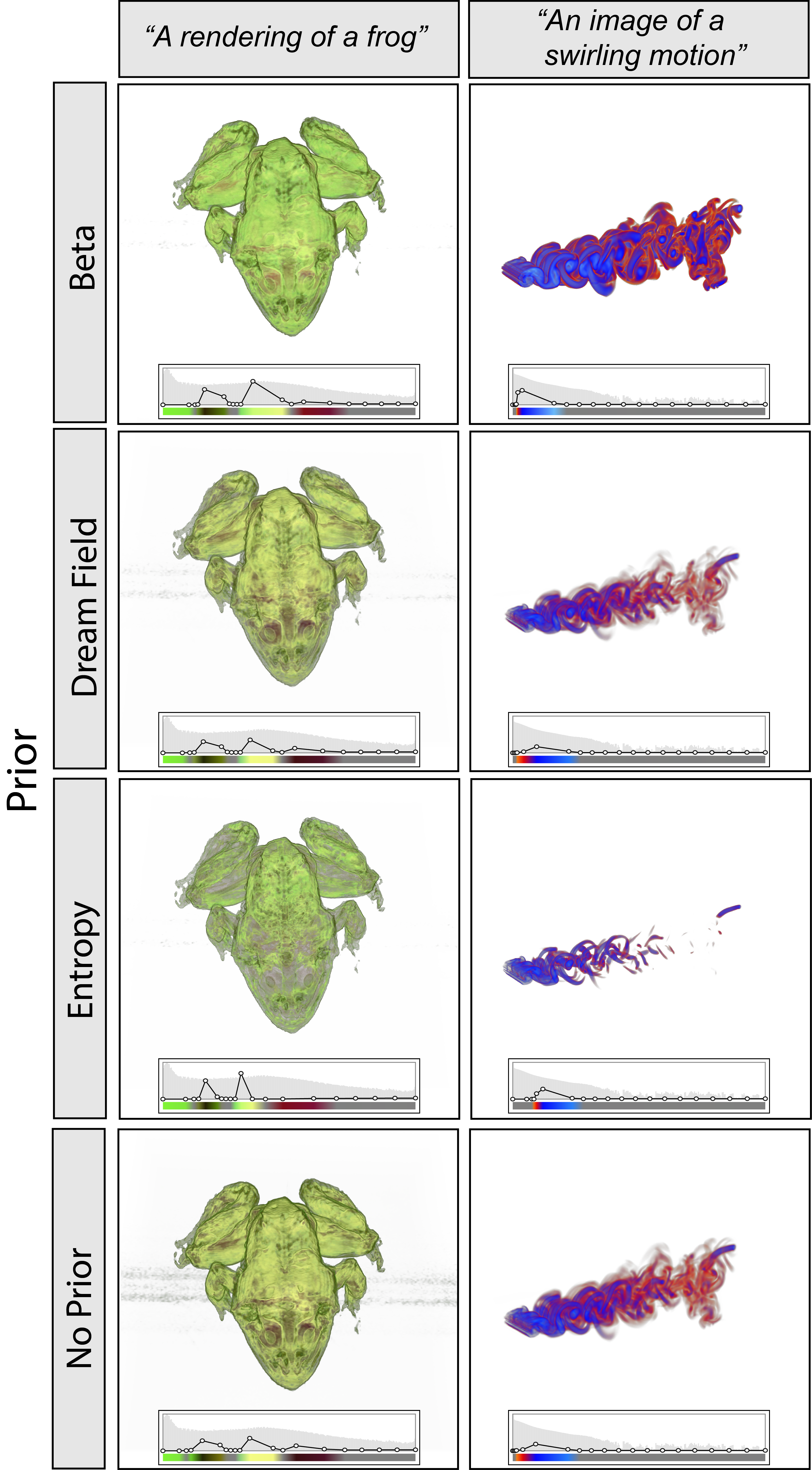}
    \caption{We observe that Beta prior consistently gives a clear volume rendering with correct opacity and fewer visual artifacts. Optimizing without a density prior often leads to noisy ambience and high transparency. Density prior from Dream Field or Entropy prior yields more transparent volume or fails to optimize.}
    \label{fig:prior-comparison}
\end{figure}

We apply density prior regularization to strengthen the material in the volume so that important materials appear less transparent in the rendered image. Existing research introduces density priors per their objectives. Lombardi et al.~\cite{Lombardi:2019} introduces Beta prior in neural volume rendering, followed by prior on mean transmittance (Dream Field) and entropy-based prior (Entropy) ~\cite{jain2022zero,kim2022infonerf}. We compare three prior density methods in Fig.~\ref{fig:prior-comparison}. Entropy and Beta produce opacity functions with higher peaks compared to the training without prior, whereas the changes in Dream Field are unclear. We notice that with Entropy, the images tend to lose materials. For example, the right column shows the flow losing its tail with Entropy. Beta density prior consistently produces images with more solid material and less noise throughout our experiment.

\subsection{Weight on Density Prior}
\begin{figure}[!t]
    \centering
    \includegraphics[width=\columnwidth]{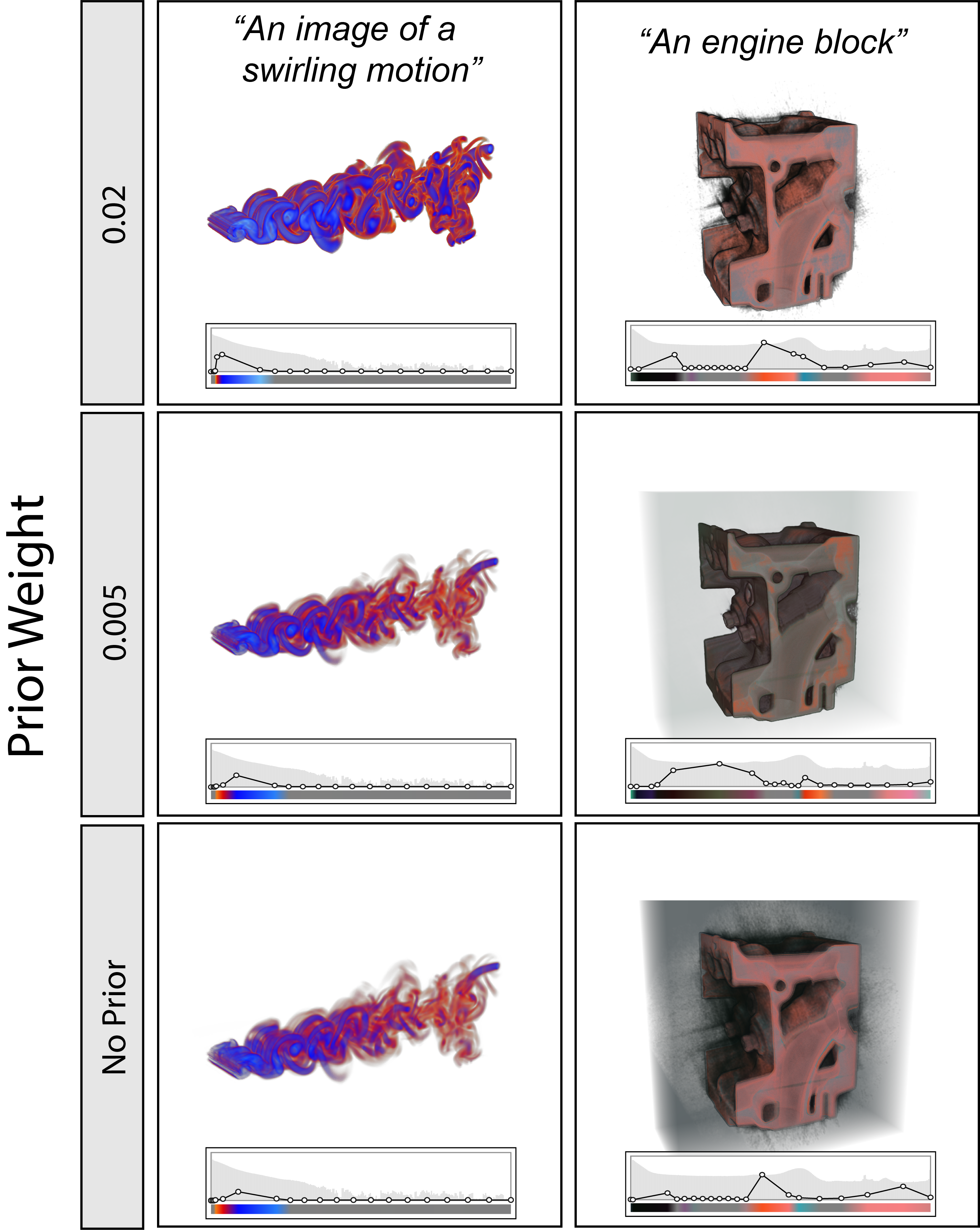}
    \caption{Choosing a correct weight on density prior is important for obtaining TF that generates a high-quality rendering. In this example, we show few cases that setting prior weight to $0.02$ leads to better rendering.}
    \label{fig:weight-comparison}
\end{figure}

We must choose a proper weight for the density prior to achieve the desired visual effect. If it is too small, the materials still look transparent, and the scene could look noisy (c.f. Fig.~\ref{fig:weight-comparison} the second row). Many of the datasets in our experiments work well with the weight of $0.02$ (c.f. Fig.~\ref{fig:weight-comparison} the first row). However, it is not necessarily optimal for other datasets. For example, the ``fossil" image in Fig.~\ref{fig:gallery} was produced with the weight of $0.001$. The optimal value is data-dependent. We hope to find an optimal way to set this parameter in future work.

\subsection{TF Regularization}

\begin{figure}[!t]
    \centering
    \includegraphics[width=\columnwidth]{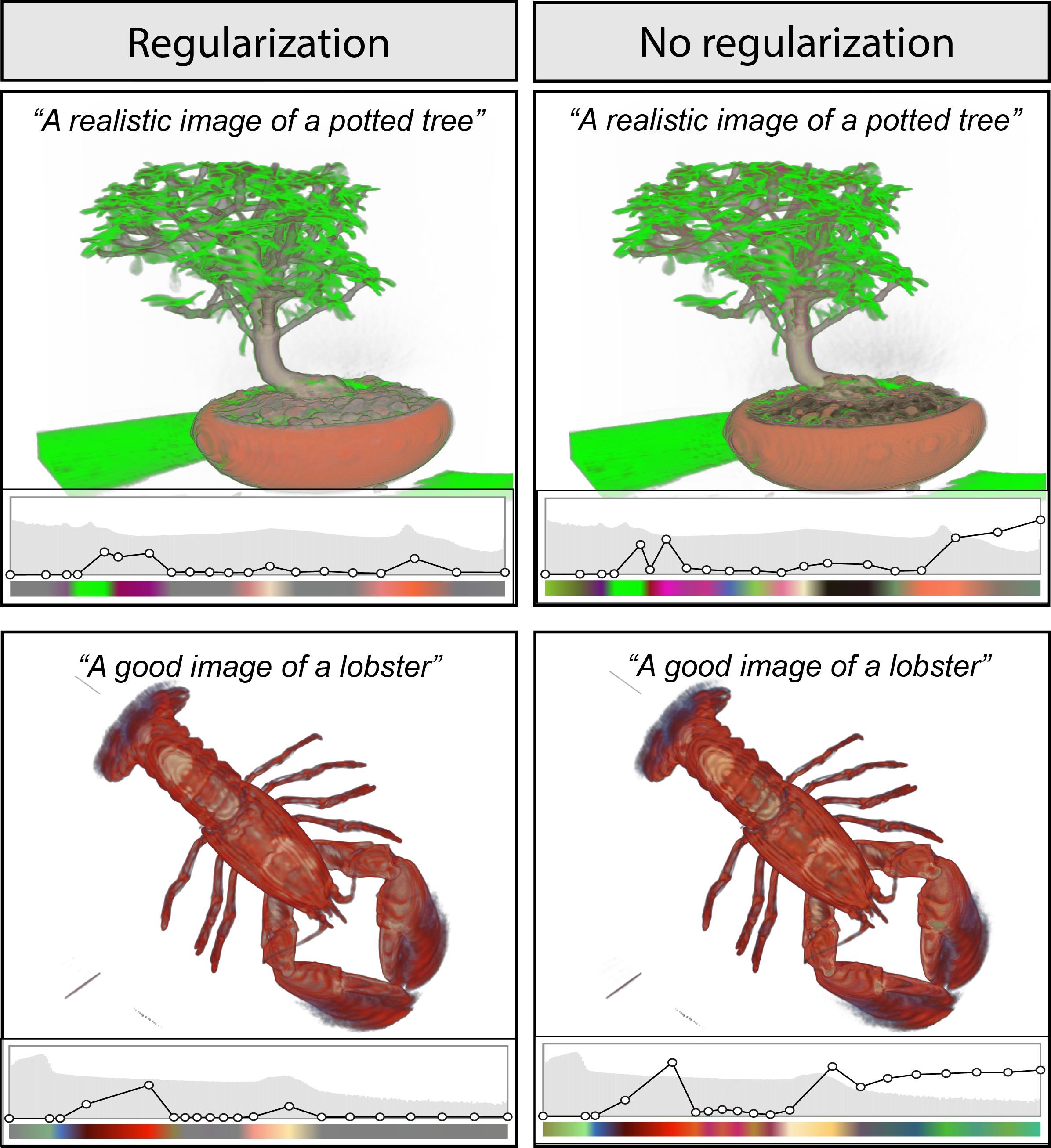}
    \caption{Imposing a regularization term on opacity and color transfer functions (TFs) simplifies them, making it easier for users to understand. Without regularization, the optimization may yield complex TFs with multiple opacity peaks and unnecessary, extra color mappings. Regularization promotes simpler, more intuitive TFs.}
    \label{fig:reg-comparison}
\end{figure}

Lastly, we find that the optimization process often times arrives at a complex TF in the presence of a potentially simpler TF that yields a similar result. To avoid unnecessary complexity, we regularize both opacity and color TF using L1 norm. In Figure \ref{fig:reg-comparison}, the effect of imposing a regularization term leads to simpler transfer functions for both opacity and color.

\newpage
\hspace{40pt}

\vspace{-10pt}

\end{document}